\documentclass[prb,twocolumn,showpacs,preprintnumbers,amssymb]{revtex4}
\usepackage{graphicx}% Include figure files
\usepackage[tbtags]{amsmath}
\usepackage{bm}% bold math

\begin{document}

\title
{Quantum signal transmission through a single-qubit chain}
\author{Ya.~S.~ Greenberg$^{1}$, C.~Merrigan$^{2}$, A.~Tayebi$^{3,4}$, and V.~Zelevinsky$^{4,5}$}

\affiliation{$^{1}$Department of Physics and Techniques,
Novosibirsk State Technical University, Novosibirsk 630092,
Russia}

\affiliation{$^{2}$William Jewell College, Liberty, Missouri
64068, USA}

\affiliation{$^{3}$Department of Electrical and Computer Engineering, College
 of Engineering, Michigan State University, East Lansing, Michigan 48824, USA}

\affiliation{$^{4}$Department of Physics and Astronomy, Michigan
State University, East Lansing, Michigan 48824, USA}

\affiliation{$^{5}$National Superconducting Cyclotron Laboratory,
Michigan State University, East Lansing, Michigan 48824, USA}

\date{\today}

\begin{abstract}

A system of a two-level atom of an impurity (qubit) inserted into
a periodic chain coupled to the continuum is studied with the use
of the effective non-Hermitian Hamiltonian. Exact solutions are
derived for the quasistationary eigenstates, their complex
energies, and transport properties. Due to the presence of the
qubit, two long-lived states corresponding to the ground and
excited states of the qubit emerge outside the Bloch energy band.
These states remain essentially localized at the qubit even in the
limit of sufficiently strong coupling between the chain and the
environment when the super-radiant states are formed. The
transmission through the chain is studied as a function of the
continuum coupling strength and the chain-qubit coupling; the
perfect resonance transmission takes place through isolated
resonances at weak and strong continuum coupling, while the
transmission is lowered in the intermediate regime.

\end{abstract}
\maketitle
\section{Introduction}

Open quantum systems are currently in the center of attention of physicists in different
subfields. The main driving force in this direction is evidently the quest for the new
progress of quantum informatics. Another area with significant recent achievements related
to openness of a quantum system is nuclear physics, where the understanding of structure
and reactions of loosely bound nuclei far from stability requires the correct unified treatment
of bound states and continuum. Cold atoms in traps and optical lattices can give rise to new effects
of coupling, entanglement and transfer of information. Solid-state micro- and nano-devices,
including Josephson junctions and spintronics, are probably the most developed arrangements
of this type.

From a general point of view, in all cases we have to deal with a mesoscopic system of interacting
constituents that serves as a guide for the transmission of a quantum signal. The system can have
intrinsic degrees of freedom which can be excited and deexcited by the signal. The coupling to
the external world is realized through a certain amount of channels characterized by the asymptotic
quantum numbers of emitted particles or quanta and the final state of the system. Each channel
has an energy threshold where it becomes open and connects to the environment. In the absence
of decoherence through external noise or a heat bath, the transmission at given energy is described
by the unitary scattering matrix in the space of channels open at this energy. All these features
are common for numerous loosely bound or marginally stable mesoscopic systems determining their main
observable properties.

A convenient mathematical formalism for description of such
systems is given by the effective non-Hermitian Hamiltonian; this
method based on the Feshbach projection formalism
\cite{Feshbach,MW} is formally exact but very flexible and can be
adjusted to many specific situations; see the recent review
article \cite{AZ11} that covers applications to the continuum
shell model in nuclear physics and, in less detail, to the quantum
signal transmission through simple periodic and disordered chains.
To stress the breadth of possible problems solved in a similar
approach we can mention recent applications to arrays of antennas
\cite{zhang12}, fullerenes \cite{sasada05} and studies of light
harvesting bacteria \cite{Cel-bact}.

One bright phenomenon emerging in the situation with a
sufficiently strong continuum coupling in the case when the number
of open channels is relatively small compared to the number of
involved intrinsic states is the so-called {\sl super-radiance}.
Being an analog of super-radiance in quantum optics
\cite{dicke54,benedict96}, this term stands for the formation in
the system of a collective superposition of the intrinsic states
coherently coupled to the same decay channel. The number of
possible states of this type is equal to the number of open
channels.

The simplest transmission system fully studied with the aid of the
effective non-Hermitian Hamiltonian is an open periodic chain with
hopping between adjacent cells
\cite{SZAnn,VZOpt,VZ777,Sorathia09,Celardo10,Sorathia12}. It can
be shown \cite{celardo09} that, after appropriate identification
of parameters, this system is physically equivalent to a realistic
sequence of quantum barriers  and to the system of parallel
waveguides considered in Ref. \cite{longhi06}. The transmission
was studied at weak continuum coupling (a system of narrow
individual resonances) and in the limit of strong coupling when
the resonances overlap and the main role is played by the physics
of super-radiance (collectivization through continuum). The limit
of overlapping resonances is usually described, in the tradition
borrowed from nuclear physics, in terms of Ericson fluctuations
\cite{ericsonAP63}; it was shown that this theory should be
corrected in a number of aspects, including the correct account
for the unitarity that is the main source of super-radiance. The
behavior of an ideal periodic system was juxtaposed to the case of
disorder, where the super-radiance still survives, and it is
possible to establish the connection to the Anderson model, degree
of chaos inside the system and universal conductance fluctuations.
The consideration was also extended to grids with two- and
three-dimensional geometry; an especially interesting case is
presented by the {\sl star graph} \cite{Ziletti} where a number of
open channels intersect at a common central point. At this point
an analog of a bound state (evanescent wave) with a long lifetime
exists that can serve for accumulation of quantum information.

Thanks to the development of nanotechnology, the preparation of
low-dimensional assemblies of nanoparticles becomes a routine
experimental task \cite{tang05}. This renews the attention to the
study of quantum properties of relatively simple low-dimensional
mesoscopic systems which reveal a rich physical behavior. In this
context, the exactly solvable models as tight-binding one- or
two-dimensional chains coupled to adatoms or impurities are of
special interest \cite{longhi06,nakamura07,dente08,garmon13}.

In what follows we consider a similar system with a two-level atom
of an impurity (qubit) inserted into an open periodic chain. The
model is simple enough to allow for the exact solution; at the
same time the model turns out to be rich enough to demonstrate
interesting physics of the signal transmission. An analytical
consideration is supplemented by the detailed numerical study for
the finite chains.

The paper is organized as follows. In Sec. 2 we consider a {\sl closed} chain of $2N$ identical cells with the nearest neighbor
hopping interaction. Two arms of the chain are connected through the central cell occupied by a two-level atom (qubit).
We find the energy spectrum of the system that consists of a normal band of delocalized Bloch standing waves and two additional
states outside the band corresponding to the excited and the ground states of the qubit. We study the evolution of
the energy spectrum as a function of the chain-qubit coupling strength.

Sec. 3 describes the energy spectrum of the {\sl open} system coupled to the continuum through its edge states.
The former stationary states acquire decay widths that change as a function of the coupling constants. The decay
widths of the qubit states are small compared to those of the Bloch waves. The qubit states remain essentially
localized at the qubit even when the coupling of the chain to the continuum becomes sufficiently strong to allow
for the formation of super-radiant states.

In Sec. 4 we study the transmission through the chain as a function of the coupling parameters. In both limits
of the edge continuum coupling being weak and strong compared to the coupling between the qubit levels, the resonances are
well isolated all having narrow decay widths and perfect transmission at resonance energies. When both couplings
are of comparable strength, the resonances are overlapped and the transmission is below the perfect level.
The results are summarized in the Conclusion.

\section{Closed chain}

\subsection{Symmetric and antisymmetric modes}

We consider a linear chain of $2N$ identical cells numbered as $n=-N,-(N-1),...,-1$ and $n=1,...,N-1,N$,
while the central cell, $n=0$, is occupied by a qubit, a two-level atom with states $|0\rangle$ and $|e\rangle$,
excitation energy $\epsilon_{e}-\epsilon_{0}=\Delta$, and matrix element $\lambda$ of the qubit excitation.
The energy $\epsilon_{0}$ is the level position in all cells; for simplicity we put the lower level of the isolated qubit at the same
position. Introducing the hopping (tunneling) matrix element $v$ between the neighboring cells, we come to the
Hamiltonian of the closed chain:
\begin{equation}
H_{nn}=\epsilon_{0}, \quad H_{n,n+1}=H_{n+1,n}=v , \quad n=-N,...,0,...,N;                    \label{1}
\end{equation}
\begin{equation}
H_{ee}=\Delta, \quad H_{0e}=H_{0e}=\lambda,                                    \label{2}
\end{equation}
where all matrix elements can be considered as real.

Before introducing the coupling to the outside world, we briefly characterize the solution for the closed chain.
A general stationary state $|E\rangle$ with energy $E$ can be presented as a superposition
\begin{equation}
|E\rangle=\sum_{n=-N}^{N}c_{n}(E)|n\rangle + b(E)|e\rangle.                 \label{3}
\end{equation}
The boundary conditions for the chain closed at the edges are $c_{-N-1}=c_{N+1}=0$.
The coefficients of the superposition (\ref{3}) satisfy the obvious equations:
\begin{equation}
(E-\epsilon_{0})c_{n}-v(c_{n-1}+c_{n+1})=0, \quad n\neq 0,                 \label{4}
\end{equation}
\begin{equation}
(E-\epsilon_{0})c_{0}-v(c_{-1}+c_{+1})=\lambda b,                  \label{5}
\end{equation}
\begin{equation}
(E-\Delta)b=\lambda c_{0}.                                         \label{6}
\end{equation}
Eqs. (\ref{5}) and (\ref{6}) can be treated as boundary conditions for the two chains (left and right)
implying that
\begin{equation}
c_{0}(E)=\,\frac{v(E-\Delta)}{(E-\epsilon_{0})(E-\Delta)-\lambda^{2}}\,(c_{-1}+c_{1}).    \label{7}
\end{equation}

Following the same procedure as for a single chain \cite{SZAnn}, see also Appendix in Ref. \cite{Celardo10},
we find that the solutions on both sides of the qubit have the form
\begin{equation}
c_{n}=\left\{\begin{array}{c}
A\xi_{+}^{n}+B\xi_{-}^{n}, \; n<0;\\
A'\xi_{+}^{n}+B'\xi_{-}^{n}, \; n>0. \end{array}\right.         \label{8}
\end{equation}
The amplitudes $\xi_{\pm}$ are given by
\begin{equation}
\xi_{\pm}=\,\frac{1}{2v}\,\left[E-\epsilon_{0}\pm\sqrt{(E-\epsilon_{0})^{2}-4v^{2}}\right], \label{9}
\end{equation}
and $\xi_{+}\xi_{-}=1$. The edge conditions determine
\begin{equation}
B=-A\,\left(\,\frac{\xi_{-}}{\xi_{+}}\,\right)^{N+1}, \quad B'=-A'\,\left(\,
\frac{\xi_{+}}{\xi_{-}}\,\right)^{N+1}.                           \label{10}
\end{equation}

The central point determines two classes of solutions. The {\sl
antisymmetric} states, $c_{n}+c_{-n}=0$, have $c_{0}=0$, so that
these wave functions are decoupled from the excited qubit that
lives on the excited level $E=\Delta$ (in this situation the qubit
can be excited or deexcited only by an additional external
coupling). Because of $c_{0}=0$ and interaction only between the
neighboring cells, the two sides of the chain are also decoupled
from each other so that such modes do not take part in the
transport through the whole chain. The spectrum $E_{q}$ of the
antisymmetric states is determined by the roots of
$\xi_{-}^{2N+2}=1$ that can be parameterized by the even-number
quantized quasimomentum $q$, or by the phase $\varphi_{q}$,
\begin{equation}
E_{q}=2v\cos\varphi_{q}, \quad \varphi_{q}=\,\frac{\pi q}{2N+2}, \quad q\;{\rm even}, \label{11}
\end{equation}
and the local amplitudes (\ref{8}) of the wave functions are
\begin{equation}
c_{n}(q)=i^{q}\,\sqrt{\,\frac{1}{N+1}\,}\sin(n\varphi_{q}).                      \label{12}
\end{equation}
The energies (\ref{11}) are inside the Bloch band $(-2v,+2v)$.

The {\sl symmetric} solutions, $c_{1}=c_{-1}$, involve the qubit into dynamics and the corresponding
roots are determined by
\begin{equation}
\xi_{+}^{2N+2}=\,\frac{\lambda^{2}+(E-\Delta)\sqrt{(E-\epsilon_{0})^{2}-4v^{2}}}{\lambda^{2}-
(E-\Delta)\sqrt{(E-\epsilon_{0})^{2}-4v^{2}}}.                       \label{13}
\end{equation}
To derive this relation, it is convenient to use eq. (\ref{4}) and connect the amplitude $c_{1}$, or $c_{-1}$,
to the central amplitude $c_{0}$,
\begin{equation}
c_{1}=c_{0}\,\frac{\xi_{+}^{N}-\xi_{-}^{N}}{\xi_{+}^{N+1}-\xi_{-}^{N+1}}.
\label{14}
\end{equation}
With the qubit disconnected, $\lambda\rightarrow 0$, we have the
missing in eq. (\ref{11}) $q$-odd part of the band spectrum with
the amplitudes (\ref{12}) where $\sin(n\varphi_{q})$ is changed to
$\cos(n\varphi_{q})$. {Depending on the relative position of
$\Delta$ with respect to the band width $2v$, there are two cases;
below throughout the paper we set $\epsilon_{0}=0$.}

%\textbf{Next we consider the case when qubit excitation level is
%outside the band.}

\subsection{Case $\Delta>2v$}

\begin{figure}
%\begin{center}
%\includegraphics[width=.95\textwidth]{fig1.eps}
\includegraphics[width=\columnwidth]{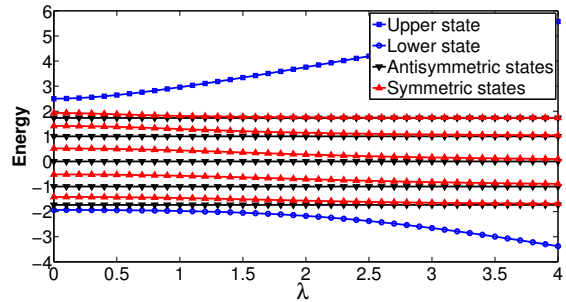}
%\end{center}
\caption{(Color online). \small{Energy levels for a system of the
closed chain of eleven sites and the excited qubit state in the
middle as a function of the qubit coupling strength $\lambda$; the
hopping amplitude is set to $v=1$, and the excitation energy of
the qubit $\Delta=2.5$.}} \label{fig:001}
\end{figure}

{In this case, the upper level is always} outside the band,
$E>2v$; the lower level also leaves the band at a finite value of
$\lambda$ when $E<-2v$. The full spectrum of twelve energy levels
for $N=5$ is shown as a function of $\lambda$ in
Fig.\ref{fig:001}. The upper state starts at $E=\Delta$ and grows
approximately linearly with $\lambda$. At small $\lambda$, this is
the state with the excited qubit and only weak admixtures of
intrinsic sites.

\begin{figure}
%\begin{center}
%\includegraphics[width=.95\textwidth]{fig2}
\includegraphics[width=\columnwidth]{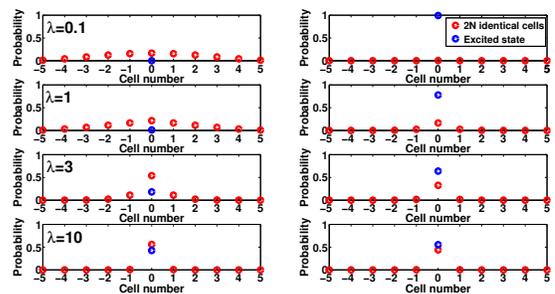}
%\end{center}
\caption{(Color online). \small{Squared components of the lower
(left column) and upper (right column) eigenstates as a
function of the qubit excitation strength $\lambda$, for a chain
$N=5$ and $v=1$.}} \label{fig:002}
\end{figure}

With increasing $\lambda$, the symmetric and antisymmetric states
inside the band become degenerate, while the ``excited" (upper) wave
function is spread almost equally over two states of the qubit.
The orthogonal combination of the excited and ground state of the
qubit gives the lower state (at sufficiently large values of
$\lambda$). In this limit the picture effectively is of Rabi
oscillations between the qubit levels with only a small
probability of hopping along the chain. The chain then is almost
decoupled from the qubit. This evolution of the two wave functions
is illustrated by Fig. \ref{fig:002}.

\subsection{Case $\Delta<2v$}

\begin{figure}
%\begin{center}
%\includegraphics[width=.95\textwidth]{fig3}
\includegraphics[width=\columnwidth]{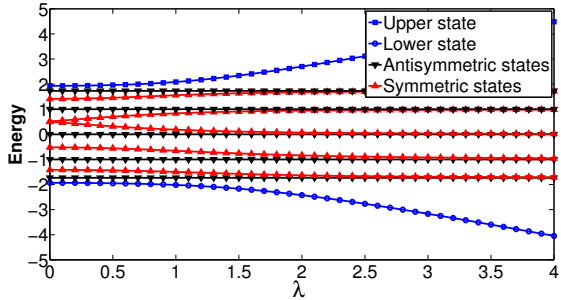}
%\end{center}
\caption{(Color online). \small{Energy levels for a system of the
closed chain of eleven sites and the excited qubit state as a
function of the qubit coupling strength $\lambda$; the hopping
amplitude is set to $v=1$, and the excitation energy of the qubit
$\Delta=0.5$.}} \label{fig:003}
\end{figure}

{Here we consider the spectrum of the energy levels as a function
of $\lambda$ when the qubit excitation energy  is inside the  band
($\Delta<2v$). The full spectrum of twelve energy levels for
$N=5$, $\Delta=0.5$ is shown as a function of $\lambda$ in Fig.
\ref{fig:003}. When $\lambda=0$, all the energies are inside the
band; with increasing $\lambda$, both the upper and lower energies
move out of the band (as before). As $\lambda$ increases the
symmetric states and the qubit level inside the band merge with
corresponding antisymmetric states.}

\begin{figure}
%\begin{center}
%\includegraphics[width=.95\textwidth]{fig4}
\includegraphics[width=\columnwidth]{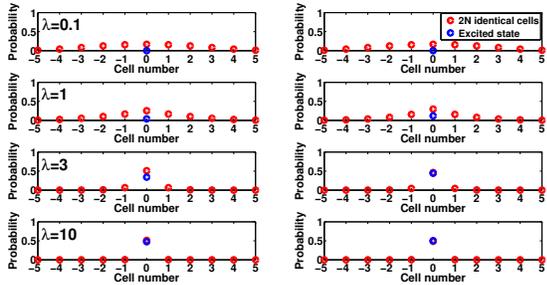}
%\end{center}
\caption{(Color online). \small{Squared components of the lower
(left column) and upper (right column) of the wave functions as a
function of the qubit excitation strength $\lambda$, for a chain
$N=5$, $\Delta=0.5$ and $v=1$.}} \label{fig:004}
\end{figure}

{Figure \ref{fig:004} shows the eigenstates corresponding to the
lowest and highest eigenenergies, upper and lower wave functions.
Unlike in the  case B, when  $\lambda= 0.1$, in the upper wave
function  there is a small admixture of the excited qubit state. }

\begin{figure}
%\begin{center}
%\includegraphics[width=.95\textwidth]{fig5}
\includegraphics[width=\columnwidth]{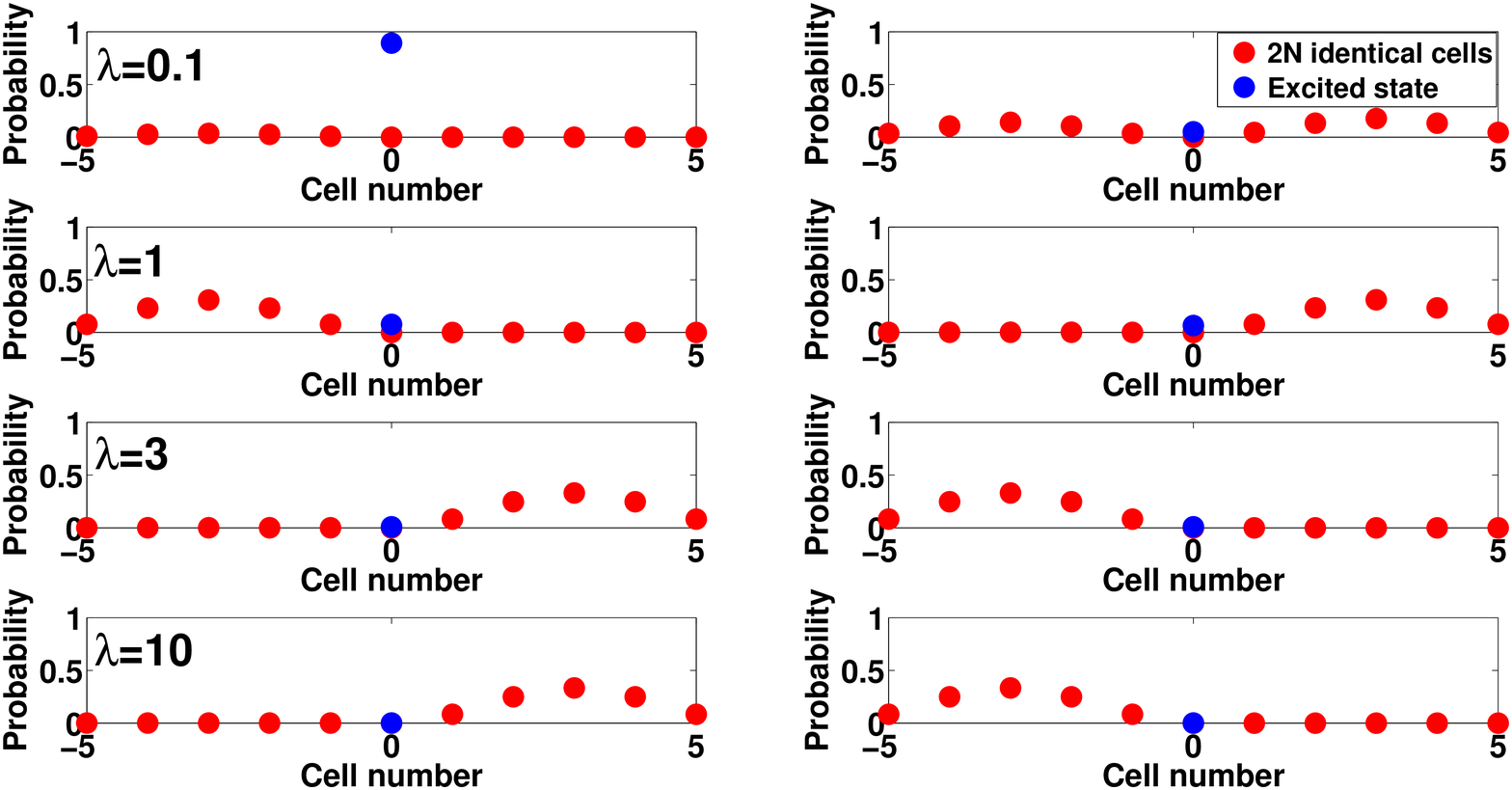}
%\end{center}
\caption{(Color online). \small{Squared components  of two wave
functions for degenerate level as a function of the qubit
excitation strength $\lambda$, for a chain $N=5$,
$\Delta=\sqrt{}3$ and $v=1$.}} \label{fig:005}
\end{figure}

{The special situation emerges if $\Delta$ is degenerate with the
energy of one of the antisymmetric states.
%These two levels are not function of
%$\lambda$ and appear to be degenerate.
In this case, $c_0=0$,
however $b$ is not necessarily vanishing. Eq. (5) results
in $c_1+c_{-1}=-(\lambda/v)b$. As a consequence we obtain
here two eigenstates which do not fall into the symmetric
or anti-symmetric category. In Fig. \ref{fig:005} we show the two
eigenstates for $N=5$, $v=1$, $\Delta=\sqrt{3}$. As
$\lambda$ increases the two state become localized either in the
left chain or in the right chain.}

\subsection{Large N chain}

{For two qubit levels outside of the band we can derive in case of
large $N$ a simple expression. For the upper qubit level, $E>2v$,
we always have $\xi_+>\xi_-$. Hence, for large $N$ we may neglect
$\xi_-^N$ in eq. (\ref{14}) to obtain $c_1=c_0/\xi_+=c_0\xi_-$.
Accordingly, for the lower qubit level, $E<-2v$, we may neglect
$\xi_+^N$ in eq. (\ref{14}). This allows us to find from
expressions (\ref{5}) and (\ref{6}) equations for the two levels
genetically related to the qubit:}
\begin{equation}\label{lgN}
    \sqrt{E^2-4v^2}=\pm\frac{\lambda^2}{E-\Delta},
\end{equation}
{where plus (minus) corresponds to the upper (lower) qubit level.}

{The wave functions of these states are given by}

\begin{equation}\label{wf}
    |\Psi _ {\mp }\rangle = {B_ \mp }\left\{\sum\limits_{ - N}^N {\xi _ \mp ^{\left| n \right|}\left| n \right\rangle }
      + \frac{\lambda }{{E - \Delta }}\left| e \right\rangle\right\},
\end{equation}
{where the normalization constant $B_\mp$ is defined by}
\begin{equation}\label{norm}
    B_ \mp ^2\left[ {1 + \frac{{{\lambda ^2}}}{{{{(E - \Delta )}^2}}}
     - 4\frac{{\varepsilon _ \mp ^2}}{{\varepsilon _ \mp ^2 - 1}}} \right] = 1,
\end{equation}
{with the same identification of the signs.}
%In eqs. (\ref{wf}) and (\ref{norm}) the upper sign $+$ corresponds to the
%ground qubit state, $\Delta<-2v$, the lower sign $-$ corresponds
%to the excited qubit state, $\Delta>2v$.}

{In fact, the asymptotic expression (\ref{lgN}) is good even for
$N=5$. For example, for $E/2v=1.1$ we have the ratio
$(\xi_-/\xi_+)^5=0.01$, while at $N=10$ the ratio
$(\xi_-/\xi_+)^{10}$ is of the order of $10^{-4}$. Therefore, for
the qubit levels positioned out of the band, the expression
(\ref{lgN}) is good for any $N\geq 5$ and for any value of
$\Delta$.} {As before, these two states move out of the band as
$\lambda$ increases. For the states inside the band, there are no
simple expressions for large $N$. However, their behavior is
similar to that shown in Fig.\ref{fig:001}: as $\lambda$ increases
all symmetric states merge with antisymmetric states.}

\section{Open chain}

Now we assume that the edge states are coupled to the outside world by ideal leads and the situation
becomes identical with what can be described by the effective non-Hermitian Hamiltonian,
\begin{equation}
{\cal H}=H-\,\frac{i}{2}W,                           \label{15}
\end{equation}
acting only in the intrinsic space of the closed system. Here the anti-Hermitian part $W$ is factorized
\cite{AZ11} having the matrix elements $W_{12}$ between the intrinsic states $|1\rangle$ and $|2\rangle$
in the form of the product of partial amplitudes $A_{1}^{c}$ and {\bf $A_{2}^{c}$} which generate the interaction between
intrinsic states through an open decay channel $c$,
\begin{equation}
W_{12}=\sum_{c;\,{\rm open}}A_{1}^{c}A_{2}^{c}.                 \label{16}
\end{equation}
It is important that here the superscript $c$ runs only over open
channels ({\sl on-shell} interaction). In general the amplitudes
$A_{1}^{c}$, {which represent the coupling amplitudes between an
intrinsic state $|1\rangle$ and channel $c$, are energy-dependent
vanishing at the threshold energy for a given channel, and their
near-threshold behavior determines the non-exponential decay curve
in the long-time limit. In many cases this dependence can be
ignored, which we assume in this work as well. In our case, the
only non-vanishing amplitudes are $A^{L}_{-N}=\sqrt{\gamma_{L}}$
and  $A^{R}_{N}=\sqrt{\gamma_{R}}$ describing the coupling of the
left (right) edge to the left (right) decay channel.}

In the case of a linear chain we allow only left and right decays
through the edge states (it would be also interesting to study the
decoherence through the coupling with many ``random" weak channels
connected to the intrinsic sites). The factorized nature of the
operator $W$ (dictated essentially by requirements of unitarity of
the scattering matrix in the channel space, see for example
\cite{durand76}), shows that this operator has only few non-zero
eigenvalues their number being equal to the number of open
channels. The corresponding eigenstates {of $W$} are obviously
just the edge states directly coupled to the continuum. Therefore
it is sufficient in our scheme to introduce two complex energies,
$\epsilon_{R}$ and $\epsilon_{L}$ for the states at the edges,
where $\epsilon_{L,R}=\epsilon_{0}-(i/2)\gamma_{L,R}$. Such an
open system without a qubit was analyzed in Refs.
\cite{SZAnn,VZ777,Celardo10,celardo09}. In this approach the
states directly coupled to the continuum play the role of doorways
\cite{AZ07}, and the remaining states can get their widths (finite
lifetimes) only through their coupling to the doorways.

The typical situation for the chain without a qubit is evolving as
a function of parameters $\gamma_{L,R}$ compared to the level
spacing $D$ in the closed system. At weak coupling, every
intrinsic state becomes a resonance with a small decay width
determined by the overlap of the Bloch state with the edges; the
final width distribution has, for $\gamma_{L}=\gamma_{R}$, a
maximum in the center of the band. In the limit of strong
continuum coupling, we have a super-radiant situation when the
central state {in the spectrum} accumulates almost the entire
width while the remaining states become very long-lived (trapped).
In the case of $\gamma_{L}\neq \gamma_{R}$ there occur two
super-radiant transitions \cite{Celardo10} with the maximum of the
signal transmission in between. In the site representation, the
super-radiant states with energies in the center of the band are
concentrated at the edges \cite{SZAnn}. This picture survives also
the possible presence of disorder in the intrinsic wells \
\cite{VZ777}.

The results for the chain with the qubit are illustrated by the
series of graphs, where the chain consists of  5+1+5=11 cells with
the qubit in the middle, altogether 12 intrinsic states. Here we
diagonalize the effective Hamiltonian in the {\sl doorway
representation}: the continuum coupling occurs only at the edges
which serve as doorways, and the matrix elements of the
anti-Hermitian part of the effective Hamiltonian (\ref{15}), which
couples the states $|q\rangle$ with the outside world, are given
in the band representation by
\begin{equation}
W_{qq'}=\gamma_{L}c_{-N}(q)c_{-N}(q')+\gamma_{R}c_{N}(q)c_{N}(q').              \label{17}
\end{equation}

\begin{figure}
%\begin{center}
%\includegraphics[width=.95\textwidth]{fig6.eps}
\includegraphics[width=\columnwidth]{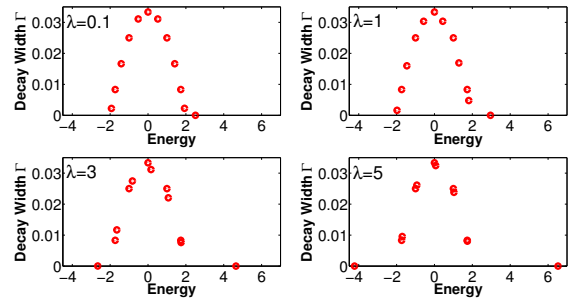}
%\end{center}
\caption{(Color online). \small{Resonance complex energies
(eigenvalues of the effective Hamiltonian for weak continuum
coupling, $\gamma=0.1$) evolve as a function of the qubit
excitation strength $\lambda$. Two qubit states are effectively
decoupled from the chain and move along the real energy axis.}}
\label{fig:006}
\end{figure}

Fig. \ref{fig:006} shows the evolution (as a function of
$\lambda$) of the resonance energies for the case of weak
continuum coupling, $\gamma_{L}=\gamma_{R}=0.1$ (the scale is
fixed by the band width, $v=1$). At small $\lambda$ {and
$\Delta>2$}, we have the parabolic distribution of widths with the
maximum at the center of the band, as known from previous studies
\cite{SZAnn,VZ777}, and the decoupled excited qubit state above
the band with zero width. As $\lambda$ increases, the symmetric
and antisymmetric states merge becoming effectively decoupled from
the qubit. The {upper} qubit state and emerging outside the band
the {lower} qubit state are still almost stationary (Rabi regime).
They are moving along the real energy axis being repelled by the
band.

\begin{figure}
%\begin{center}
%\includegraphics[width=.95\textwidth]{fig7}
\includegraphics[width=\columnwidth]{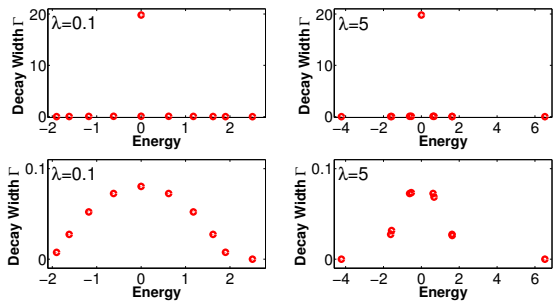}
%\end{center}
\caption{(Color online). \small{Resonance complex energies
(eigenvalues of the effective Hamiltonian for strong continuum
coupling, $\gamma=20$) show segregation of two super-radiant
states in the middle of the band from ten trapped states which
include the strongly localized states of the qubit.}}
\label{fig:007}
\end{figure}

The situation changes when we come to the strong continuum
coupling, Fig. \ref{fig:007}. Here we again follow the evolution as
a function of $\lambda$ but at $\gamma=20$. At small
$\lambda=0.1$, the two coinciding super-radiant states are formed
in the center of the band (here we keep $\gamma_{L}=\gamma_{R}$),
and the remaining ten states are trapped, effectively returning to
the non-overlap regime (the left lower plot shows, at a much
smaller width scale, the parabolic width distribution). With
increase of $\lambda$, the super-radiant states survive, while the
qubit states again are repelled by the band along the real energy
axis having still very small widths.

\begin{figure}
\begin{center}
\includegraphics[width=\columnwidth]{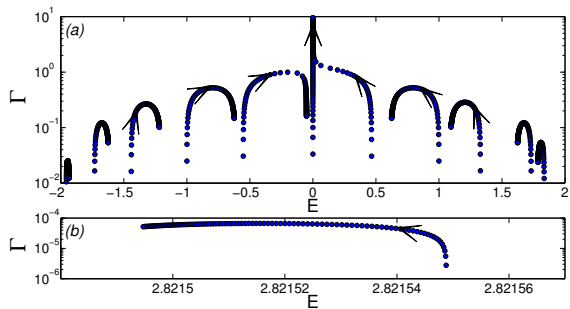}
\end{center}
\caption{(Color online). \small{Complex-plane trajectories of
eigenstates of the effective Hamiltonian; the parameter values are
$v=1,\;\lambda=0.8,\;\Delta=2.5$. The arrows show the direction of the evolution as
$\gamma$ changes from 0 to 10. Panel ({\sl a}) shows the
behavior of 12 states in the chain while panel ({\sl b}) singles
out the state genetically connected to the excited qubit state
located outside the band.}} \label{fig:008}
\end{figure}

It is instructive to take a look of the width evolution as a
function of the continuum coupling strength $\gamma$. The typical
process is presented by Fig. \ref{fig:008}, where the trajectories of all
twelve complex poles in the lower half of the complex plane are
shown. In the limit of very weak continuum coupling, part ({\sl
a}) of this figure, the width distribution is parabolic, being
proportional, as a function of real energy, to the group velocity
of the band states. It is transformed with increase of $\gamma$.
All widths, except for the super-radiant states in the center of
the energy band, turn back after reaching their maximum values.
Their corresponding trajectories are almost symmetric with respect
to their maxima which is typical for the phenomenon of
super-radiance in a space of fixed dimension. Indeed, after the
segregation of the super-radiant state(s), the remaining trapped
states are essentially in the same situation as they were in the
beginning of the process; this symmetry for the finite dimension
of intrinsic space is a characteristic feature \cite{Celardo10}
that appears also in the statistical distribution of neutron
widths for thermal-energy neutron resonances \cite{Gav12}. Part
({\sl b}) of the figure selects, on a detailed energy scale, the
complex-plane evolution of the excited qubit state that becomes
extremely long-lived. The details of interference between
neighboring resonances were also discussed repeatedly in the
context of the electron conduction in nano-scale systems, see for
example \cite{sasada05-11}.

\begin{figure}
\begin{center}
\includegraphics[width=\columnwidth]{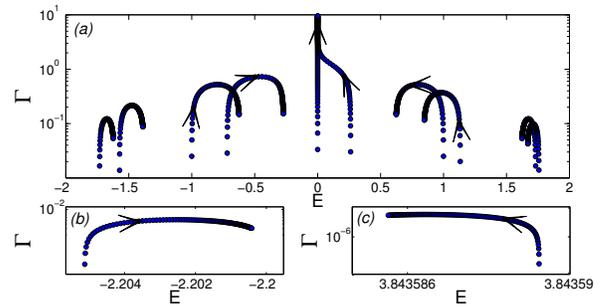}
\end{center}
\caption{(Color online). \small{Complex-plane trajectories of
eigenstates of the effective Hamiltonian; the parameters values
are $v=1,\;\lambda=2.1,\;\Delta=2.5$. Ten states, including the
super-radiant at the edges of the chain, are moved closer inside
the band, panel ({\sl a}), two states mainly localized at the
qubit have energies outside the band and essentially interact only
with each other having a large lifetime with respect to tunneling
through the chain, panels ({\sl b}) and ({\sl c})}.}
\label{fig:009}
\end{figure}

It was noticed long ago \cite{brentano96} that, in the description of an open system with the aid of
the effective Hamiltonian, the Hermitian and non-Hermitian parts of the interaction act in the opposite way.
The real (Hermitian) perturbation repels the levels but, through the mixing mechanism, attracts the widths
of unstable states. Contrary to that, the imaginary (non-Hermitian) interaction through the continuum
repels the widths (the road to superradiance and trapping) but attracts real energies of resonances.
This attraction is seen in Fig. \ref{fig:009} for a larger value of $\lambda$: as the continuum coupling $\gamma$
increases along the road to super-radiance, the real energies of the poles move to the middle of the band,
where the super-radiant states are located.

\begin{figure}
\begin{center}
\includegraphics[width=\columnwidth]{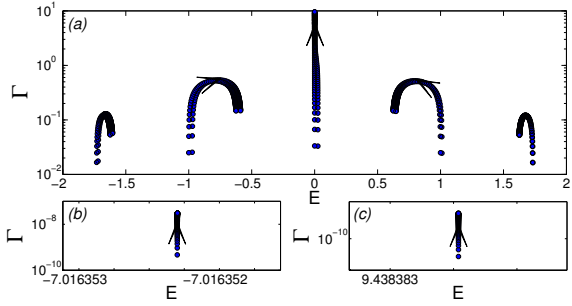}
\end{center}
\caption{(Color online). \small{Complex-plane trajectories of
eigenstates of the effective Hamiltonian; the parameters values
are $v=1,\;\lambda=8,\;\Delta=2.5$. Ten states, including the
super-radiant at the edges of the chain, are moved closer inside
the band, panel ({\sl a}). The qubit states, the ground state,
({\sl b}), and the excited state, ({\sl c}), are essentially
decoupled.}} \label{fig:010}
\end{figure}

Finally, the limit of very strong coupling between the qubit
levels is shown in Fig. \ref{fig:010} where $\lambda=8$. With increasing continuum coupling $\gamma$,
as antisymmetric and symmetric roots merge,
we see, panel ({\sl a}), the evolution of pairs of states (ten of
them inside the band including the super-radiant ones). The time
arrow of this evolution is again in the direction of attraction
for the real energies of resonances. The qubit states, the ground
state, ({\sl b}), and the excited state, ({\sl c}), are
essentially decoupled. They do not shift and only gradually
increase their (still small) decay widths as $\gamma$ increases.

\begin{figure}
\begin{center}
\includegraphics[width=\columnwidth]{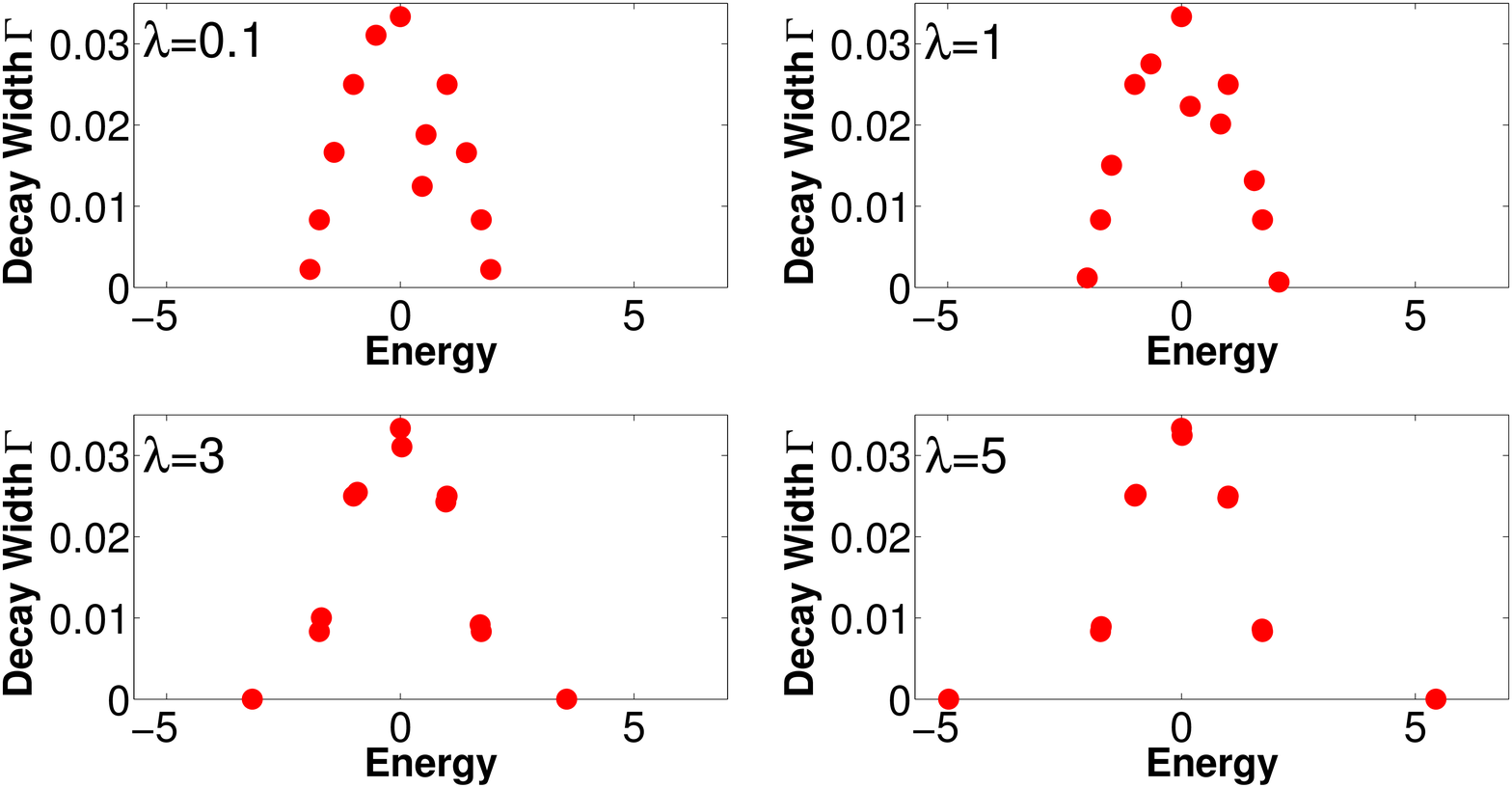}
\end{center}
\caption{(Color online). \small{Resonance complex energies
(eigenvalues of the effective Hamiltonian) for $\Delta=0.5$ and
for weak continuum coupling, $\gamma=0.1$ as a function of the
qubit excitation strength $\lambda$.}} \label{fig:011}
\end{figure}

\begin{figure}
\begin{center}
\includegraphics[width=\columnwidth]{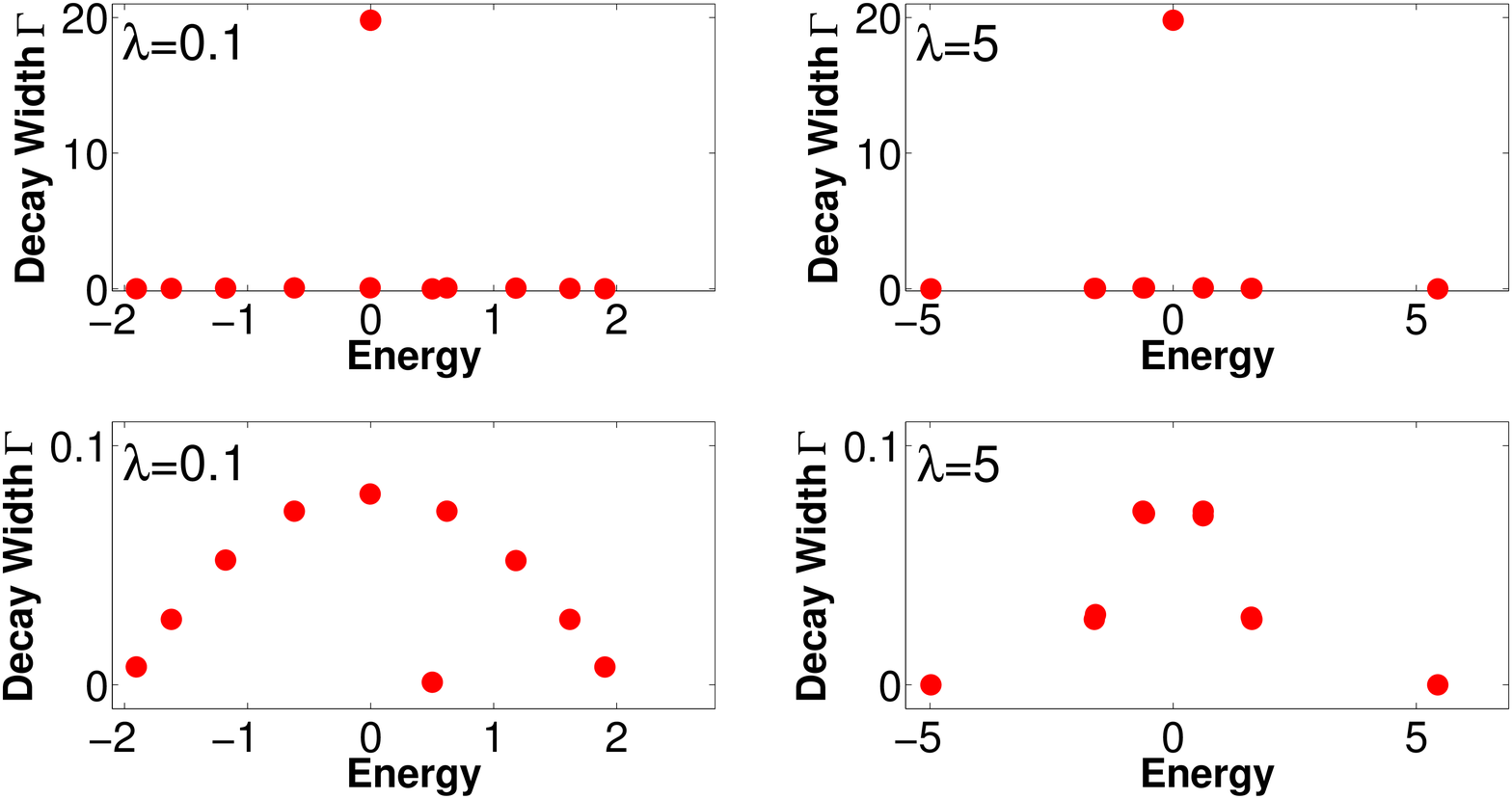}
\end{center}
\caption{(Color online). \small{Resonance complex energies
(eigenvalues of the effective Hamiltonian) for $\Delta=0.5$ and
for strong continuum coupling, $\gamma=20$ as a function of the
qubit excitation strength $\lambda$.}} \label{fig:012}
\end{figure}

{Next we briefly consider the case when qubit energy $\Delta$ is
inside the band. In Fig. \ref{fig:011} we show the resonance
complex energies for $\Delta=0.5$ and for weak continuum coupling,
$\gamma = 0.1$, as a function of the qubit excitation strength
$\lambda$. We have already seen that at small  $\lambda$ there are
two  eigenstates  with eigenenergies close to $\Delta= 0.5$. Their
decay  width  is smaller since there is  a strong contribution
from  the excited qubit state  to these two. As $\lambda$
increases, the eigenenergies inside the band merge,so that the two
qubit states are effectively decoupled from the chain moving along
the real energy axis.}
{Resonance complex energies for strong continuum coupling
$\gamma=20$ are shown in Fig. \ref{fig:012} for $\Delta=0.5$. Here
the situation is qualitatively identical with that in Fig.
\ref{fig:007}. At small $\lambda=0.1$, the two coinciding
super-radiant states are formed in the center of the band, and the
remaining ten states are trapped. With the increase of $\lambda$,
the super-radiant states survive, while the qubit states again are
repelled by the band along the real energy axis having still very
small widths.}

\section{Transmission through the chain}

Here we briefly consider the transmission of an external signal through the chain with the inserted qubit,
similarly to the consideration made earlier for the uniform chain or multi-dimensional lattices \cite{Celardo10}
and for the star graph \cite{Ziletti}. The internal propagation of the signal of given energy $E$ in the open
system is described by the propagator
\begin{equation}
G(E)=\,\frac{1}{E-{\cal H}},                           \label{18}
\end{equation}
where ${\cal H}$ is the effective Hamiltonian (\ref{15}) with the imaginary part (\ref{16}) that describes multiple
excursions of the signal into continuum and back. Our simple geometry has two open channels, left and right
(labeled in eq. (\ref{17}) $L$ and $R$). The full amplitude $Z^{ba}$ of the process between channels $a$ and $b$ starts with
the entrance amplitude $A^{a}_{n}$ that populates the intrinsic state $|n\rangle$ and ends with the exit
amplitude $A^{b\ast}_{m}$ from the intrinsic state $|m\rangle$; all paths $b\rightarrow a$ interfere:
\begin{equation}
Z^{ba}(E)=\sum_{mn}A^{b\ast}_{m}G_{mn}(E)A^{a}_{n}.             \label{19}
\end{equation}
It is easy to show that the corresponding scattering matrix, $S^{ba}=\delta^{ba}-iZ^{ba}$, is unitary
\cite{AZ11}. The transmission coefficient is given by
\begin{equation}
T^{ba}(E)=\left|Z^{ba}(E)\right|^{2}.                               \label{20}
\end{equation}

It might be convenient to perform the transformation to the (biorthogonal) basis $|r\rangle$ of eigenfunctions
of the effective Hamiltonian. The complex energies ${\cal E}_{r}=E_{r}-(i/2)\Gamma_{r}$ correspond to the poles
of the scattering matrix, while the process amplitude (\ref{19}) still has factorized residues transformed to
the eigenbasis,
\begin{equation}
Z^{ba}(E)=\sum_{r}\,\frac{\tilde{A}^{b}_{r}\tilde{A}^{a}_{r}}{E-{\cal E}_{r}}.   \label{21}
\end{equation}
We can note parenthetically that this description can be treated as a simple superposition of interfering
resonances only approximately, namely if the energy dependence of continuum amplitudes $A^{a}_{n}$ is neglected
as it is done in our consideration. In this approximation the time decay curve of a single isolated resonance
$|r\rangle$ would be pure exponential with the width $\Gamma_{r}$.

For the calculation of transmission we adopt the approach of Ref.
\cite{Celardo10}. For the open chain of Sec. 3, the transmission
is determined by the edge couplings which we again assume here to
be equal, $\gamma_{L}=\gamma_{R}=\gamma$. Similarly to Ref.
\cite{Celardo10}, the transmission coefficient (\ref{20}) can be
written as
\begin{equation}
T^{RL}(E)=T^{LR}(E)=\left|\,\frac{(\gamma/v^{2})(E-\Delta)}{\prod_{r=1}^{2N+2}[(E-{\cal E}_{r})/v]}\,\right|^{2}.
                                                                               \label{22}
\end{equation}
Below we show the transmission results for the chain of $N=5$ (twelve intrinsic states) and various combinations
of the parameters. The resulting picture is determined by the counterplay of the trend to super-radiation and
decoupling of the qubit.

\begin{figure}
\begin{center}
\includegraphics[width=\columnwidth]{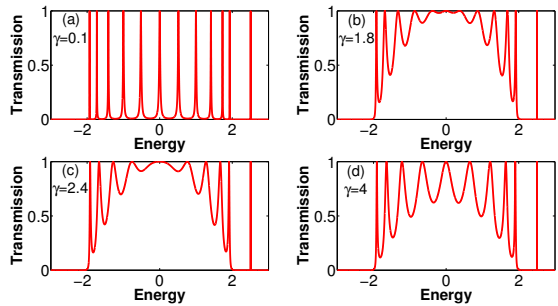}
\end{center}
\caption{(Color online). \small{Evolution of the transmission for
small $\lambda=0.1$ as a function of the continuum coupling
parameter $\gamma_{L}=\gamma_{R}=\gamma$, from 12 isolated
resonances (one outside the energy band) through overlap and
super-radiance to ten resonances corresponding to trapped states.
As earlier, $\Delta=2.5,\;v=1$.}} \label{fig:013}
\end{figure}

Starting with the weak qubit excitation amplitude, Fig. \ref{fig:013} for $\lambda=0.1$, we follow the evolution of the transmission
as a function of the continuum coupling $\gamma$. At small $\gamma=0.1$, panel ({\sl a}), we see
twelve isolated resonances all having narrow decay widths; one of them is outside the energy band as we discussed
earlier. When $\gamma$ is growing, panels ({\sl b}) and ({\sl c}), the resonances start overlapping. When two
super-radiant states merge at $\gamma=2.4$, panel ({\sl c}), they disappear from the transmission spectrum,
so that panel ({\sl d}) for $\gamma=4$ shows ten separated resonances. At each resonance the transmission
is perfect, $T=1$.

\begin{figure}
\begin{center}
\includegraphics[width=\columnwidth]{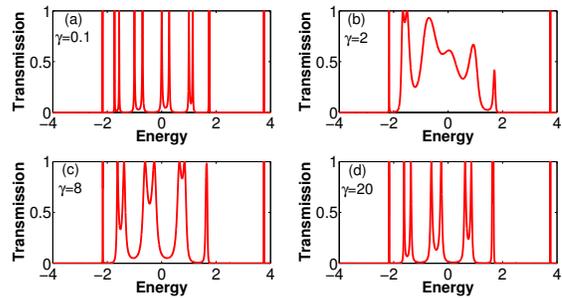}
\end{center}
\caption{(Color online). \small{The same as Fig. 13, with
$\lambda=2$.}} \label{fig:014}
\end{figure}

Next four panels, Fig. \ref{fig:014}, correspond to the intermediate value $\lambda=2$. Again the case of weak continuum
coupling, panel ({\sl a}), $\gamma=0.1$, reveals twelve resonances; now the two states associated with the qubit
are outside the energy band, while all resonances still show perfect transmission. At $\gamma=\lambda=2$,
panel ({\sl b}), the competition between the transmission through the chain and Rabi dynamics of the qubit
leads to an almost random pattern of overlapping resonances with transmission below perfect, similarly
to Ericson fluctuations \cite{ericsonAP63} or universal conductance fluctuations \cite{beenakker97}.
The relation between those well known pictures and necessary changes due to the effects of unitarity in
exact theory were discussed in Ref. \cite{Sorathia09}. At large $\gamma\gg\lambda$, panels ({\sl c})
and ({\sl d}), the continuum coupling prevails leading to the narrow resonances coming from trapped intrinsic states.

\begin{figure}
\begin{center}
\includegraphics[width=\columnwidth]{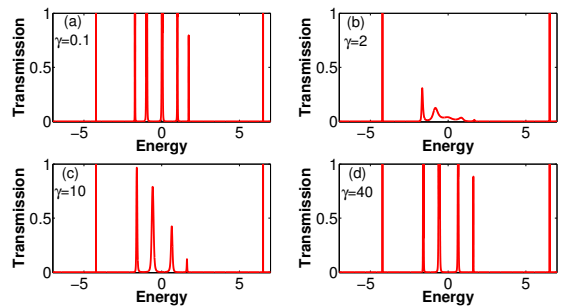}
\end{center}
\caption{(Color online). \small{The same as Fig. 13, with
$\lambda=5$.}} \label{fig:015}
\end{figure}

Finally, the large value of the qubit excitation strength,
$\lambda=5$, changes the transmission picture at not very large
$\gamma$, Fig. \ref{fig:015}. At small continuum coupling, $\gamma\ll
\lambda$, panel ({\sl a}), the two states associated with the
qubit produce two resonances outside the band, with transmission
equal to 1.  At $\gamma=2$, panel ({\sl b}), we observe some kind
of an intrinsic resonance between propagation and internal
oscillations which, along with the emergence of super-radiance,
almost kills the transmission at other energies within the band.
With further growth of $\gamma$, panels ({\sl c}) and ({\sl d}),
the perfect transmission through trapped states, including those
associated with the qubit, is gradually restored. This abundance
of possible regimes opens the way to various applications.

\begin{figure}
\begin{center}
\includegraphics[width=\columnwidth]{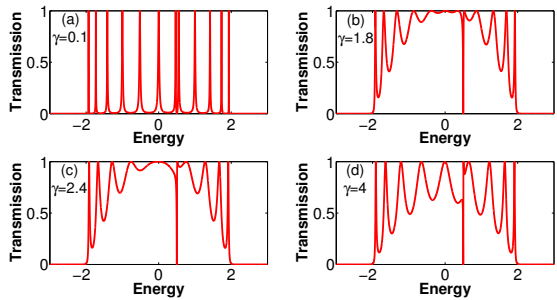}
\end{center}
\caption{(Color online). \small{Transmission through the chain for
$N=5$, $\lambda=0.1$,  $\Delta=0.5$ as function of continuum
coupling $\gamma$}.} \label{fig:016}
\end{figure}

\begin{figure}
\begin{center}
\includegraphics[width=\columnwidth]{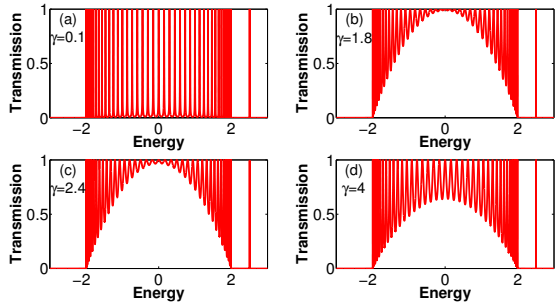}
\end{center}
\caption{(Color online). \small{Transmission through the chain for
$N=20$, $\lambda=0.1$,  $\Delta=2.5$ as function of continuum
coupling $\gamma$}.} \label{fig:017}
\end{figure}

{At the conclusion of this section we show the transmission across
the chain for the case when $\Delta$ is inside the band, Fig.
\ref{fig:016}, and for the case of large $N$, Fig. \ref{fig:017}.}

\section{Conclusion}

Using the formalism of the effective non-Hermitian Hamiltonian, we
studied a model of quantum signal transmission through a linear
periodic chain with a qubit placed at the center. For the closed
chain, the intrinsic eigenstates form two classes, antisymmetric
{that includes} standing waves with the excluded center site and
two independent subchains, and symmetric that {reveals} the qubit
dynamics with two additional eigenstates. When the system is
coupled to the environment {at its entrance and exit points}, we
have found the spectrum of quasistationary states characterized by
complex energies and finite lifetimes. Two of those states are
genetically related to the ground and excited state of the qubit
with real energy outside the Bloch band.

The most interesting feature is the stability of states related to the qubit, small decay widths and correspondingly
long lifetimes. Further, these states are only weakly perturbed when the coupling of the chain to the continuum
becomes strong. In the limit of strong coupling, the qubit states and remaining trapped Bloch states are practically
shielded from the external world by the two super-radiant states localized at the edges of the chain. The stability
of the qubit states and the possibility to switch on and off the access to them suggests that the simple configuration
considered above may serve as a building block of a quantum computer.

We discussed also the transmission through the chain as a function of the coupling strength to the continuum.
In the limits of weak and strong continuum coupling (as compared to the excitation amplitude of the qubit)
the chain reveals well separated narrow resonances with perfect transmission at corresponding energy. In the intermediate
regime, when the continuum coupling and the excitation strength of the qubit are comparable, the resonances overlap
with the transmission below the perfect level.

There are many possibilities to enrich this prototypical model.
For a realistic situation, for example a chain of quantum dots,
one should carefully determine the lifetimes of the qubit states.
The geometry of the system can be made more complicated in various
ways including the transition to more-dimensional schemes. More
qubits and more branches can be added approaching a complicated
network. It would be also interesting to extend the ideology of an
open quantum system to the study of coherent photon transport in
continuous waveguides \cite{shen05} and circuit quantum electrodynamics
\cite{zhang12,liao10, delanty11}.

\section{Acknowledgements}

Y.G. acknowledges useful discussions with A.A. Shtygashev and partial support from the Russian Ministry of Education
and Science through the project TP 7.1667.2011 and from the German Ministry of Science (BMBF) through the project
RUS 10/015. C.M. is grateful for support at MSU in the framework of the REU program.
V.Z. acknowledges the support from the NSF grant PHY-1068217.

\end{document}